\documentclass[aps,twocolumn,superscriptaddress,showpacs,amssymb,prb,floatfix,longbibliography]{revtex4-2}
\usepackage{graphicx }
\usepackage[]{hyperref}
\usepackage{dcolumn}
\hypersetup{colorlinks=true,linkcolor=blue,citecolor=blue,urlcolor=blue,pdfpagemode=UseNone}

\begin{document}

\title{Hyperfine Couplings as a Probe of Orbital Anisotropy in Heavy Fermion Materials}

\author{P. Menegasso}
\author{J. C. Souza}
\affiliation{Instituto de F\'isica \lq\lq Gleb Wataghin\rq\rq, UNICAMP, Campinas-SP, 13083-859, Brazil}

\author{I. Vinograd}
\author{Z. Wang}
\author{S. P. Edwards}
\affiliation{Department of Physics, University of California, Davis, California 95616, USA}

\author{P. G. Pagliuso}
\affiliation{Instituto de F\'isica \lq\lq Gleb Wataghin\rq\rq, UNICAMP, Campinas-SP, 13083-859, Brazil}

\author{N. J. Curro}
\affiliation{Department of Physics, University of California, Davis, California 95616, USA}

\author{R. R. Urbano}
\affiliation{Instituto de F\'isica \lq\lq Gleb Wataghin\rq\rq, UNICAMP, Campinas-SP, 13083-859, Brazil}

\begin{abstract}
The transferred hyperfine interaction between nuclear and electron spins in an heavy fermion material depends on the hybridization between the $f$-electron orbitals and those surrounding a distant nucleus.  In CeMIn$_5$ (M=Rh, Ir, Co), both the hyperfine coupling to the two indium sites as well as the crystalline electric field at the Ce are strongly dependent on the transition metal.  We measure a series of  CeRh$_{1-x}$Ir$_x$In$_5$ crystals and find that the hyperfine coupling reflects the orbital anisotropy of the ground state Ce 4$f$ wavefunction.   These findings provide direct proof that the localized to itinerant transition is dominated by hybridization out of the Ce-In plane in this system.
\end{abstract}

\maketitle

The tetragonal CeMIn$_5$ (M = Co, Rh, Ir) materials are prototypical heavy fermion systems, exhibiting quantum criticality, antiferromagnetism and  unconventional superconductivity across a phase diagram that can be tuned with pressure, magnetic field, or substitution at the transition metal site \cite{Thompson2001,tuson,tusonNature2008,YangPinesNature}. One of the outstanding mysteries in these materials is how the transition metal M changes the ground state. Similar physics is at play in the actinide PuMGa$_5$ (M=Co, Rh) materials, where the superconducting transition temperature is an order of magnitude larger \cite{PuCoGa5discovery,Wastin2003, Curro2005}.  For parts of the phase diagram where the ground state is superconducting,  $T_c$ appears to correlate with the lattice anisotropy at ambient pressure \cite{Pagliuso2002b,BauerPRL}.  However, this relationship breaks when the ground state evolves towards antiferromagnetism. Recently,  an X-ray absorption spectroscopy  (XAS) study has probed the nature of the Ce 4$f$ crystalline electrical field (CEF) ground state wavefunctions for several different CeRh$_x$Ir$_{1-x}$In$_5$ crystals \cite{Severing}. These studies revealed a strong change in the shape of the wavefunction, lending support to the idea that hybridization to the 4$f$ electron is strongly momentum-dependent \cite{Burch2007,shim2007modeling}.  Electronic structure calculations indicated that the momentum dependence affects the hybridization with the in-plane In(1) and out-of-plane In(2) 5$p$ electrons differently, leading to multiple hybridization gaps at low temperature.  Surprisingly, the In(2) appear to be more strongly coupled to the 4$f$ moments, suggesting that substitution at the M site may affect this coupling and hence the nature of the ground state.

In order to investigate the nature of this hybridization in more detail, we have investigated the NMR Knight shift in  CeRh$_x$Ir$_{1-x}$In$_5$ for both In(1) and In(2) sites.  The In nuclear spins ($I=9/2$) experience a transferred hyperfine field to the Ce 4$f$ moments that reflects the hybridization to the 4$f$ state \cite{Curro2006a,Curro2010}.  The hyperfine coupling between a nuclear spin $\mathbf{I}$ at $\mathbf{r}=0$ and an electronic spin at $\mathbf{r}$ is given by:
\begin{equation}
    \mathcal{H}_{hyp} = g\mu_B\gamma\hbar\mathbf{I}\cdot\left( \frac{8}{3}\pi\delta(\mathbf{r})+3\frac{\mathbf{r}(\mathbf{S}\cdot\mathbf{r})}{r^5}-\frac{\mathbf{S}}{r^3}\right),
\end{equation}
where $\gamma$ is the nuclear gyromagnetic ratio and $\mu_B$ is the Bohr magneton.  The first term is the Fermi contact term relevant for $s$-orbitals, and the second and third terms constitute a dipolar interaction with the electron spin.  Typically both contact and dipolar terms are present, as well as multiple electron spins, leading to an effective hyperfine coupling tensor that  in practice is determined empirically and contains both Fermi-contact and dipolar components.  Often the magnitude of the dipolar component exceeds the direct dipolar field for a localized spin by at least an order of magnitude - e.g. in this case the dipolar field of a moment located at the origin of the Ce atom. This enhancement is due to hybridization of the orbitals of the unpaired electron spin with the relevant orbitals surrounding the nucleus.  These so-called  \emph{transferred} hyperfine couplings depend sensitively on the electronic wavefunction \cite{Renold2001}.

In heavy fermions, there are two sets of electronic spins: those associated with the itinerant conduction electrons, $\mathbf{S}_c$, and with the 4$f$ orbitals, $\mathbf{S}_f$.  Because of the large spin-orbit coupling, the latter necessarily refers to the $J=5/2$ multiplet. There are different hyperfine coupling tensors to these two degrees of freedom:  $\mathcal{H}_{hyp} =  g\mu_B\gamma\hbar\mathbf{I}\cdot\left(\mathbb{A}\cdot\mathbf{S}_c + \sum_i \mathbb{B}_i\cdot\mathbf{S}_f(\mathbf{r}_i)\right)$, where $\mathbb{A}$ corresponds to an on-site coupling to the conduction electron spins \cite{ShirerPNAS2012}, and $\mathbb{B}_i$ are the transferred couplings to the nearest neighbor 4$f$ spins. The transferred couplings can be determined by comparing the Knight shift and bulk susceptibility as a function of temperature and field direction, and have been well documented for the stoichiometric CeMIn$_5$ materials \cite{ShirerPNAS2012,Curro2004,Ir115anisotropy,KambeShiftCeIrIn5}. Surprisingly, the transferred coupling $B_{cc}(1)$ for the In(1) site decreases by a factor of three between M=Rh to M=Co, whereas $B_{cc}(2)$ for the In(2) site increases by the same factor. A similar evolution of $B_{cc}(1)$ has been observed in CeRhIn$_5$ under modest hydrostatic pressure as the ground state evolves from antiferromagnetic to superconducting \cite{Lin2015}. Such a large variability in transferred hyperfine couplings constants has not been observed in other strongly correlated superconductors, such as the cuprates, or iron pnictides or chalcogenides.

We posit that in the CeMIn$_5$ materials the transferred hyperfine couplings arises due to the hybridization between the Ce 4$f$ ground state orbital and the In 5$p$ states, and that the variations in coupling constant reflect changes to the CEF parameters.  To test this hypothesis, we present a systematic analysis of the Knight shift, $K$, and bulk susceptibility, $\chi$, in a series of CeRh$_x$Ir$_{1-x}$In$_5$ crystals to extract the transferred hyperfine couplings $B_{cc}$ to the In(1) and In(2) sites as a function of $x$.  We find that these couplings provide a direct measure of the Ce orbital anisotropy and momentum-dependent hybridization.

\section*{\label{sec:experiment}Results}

\subsection*{Magnetic Susceptibility}
\begin{figure}
\centering
\includegraphics[width=\linewidth]{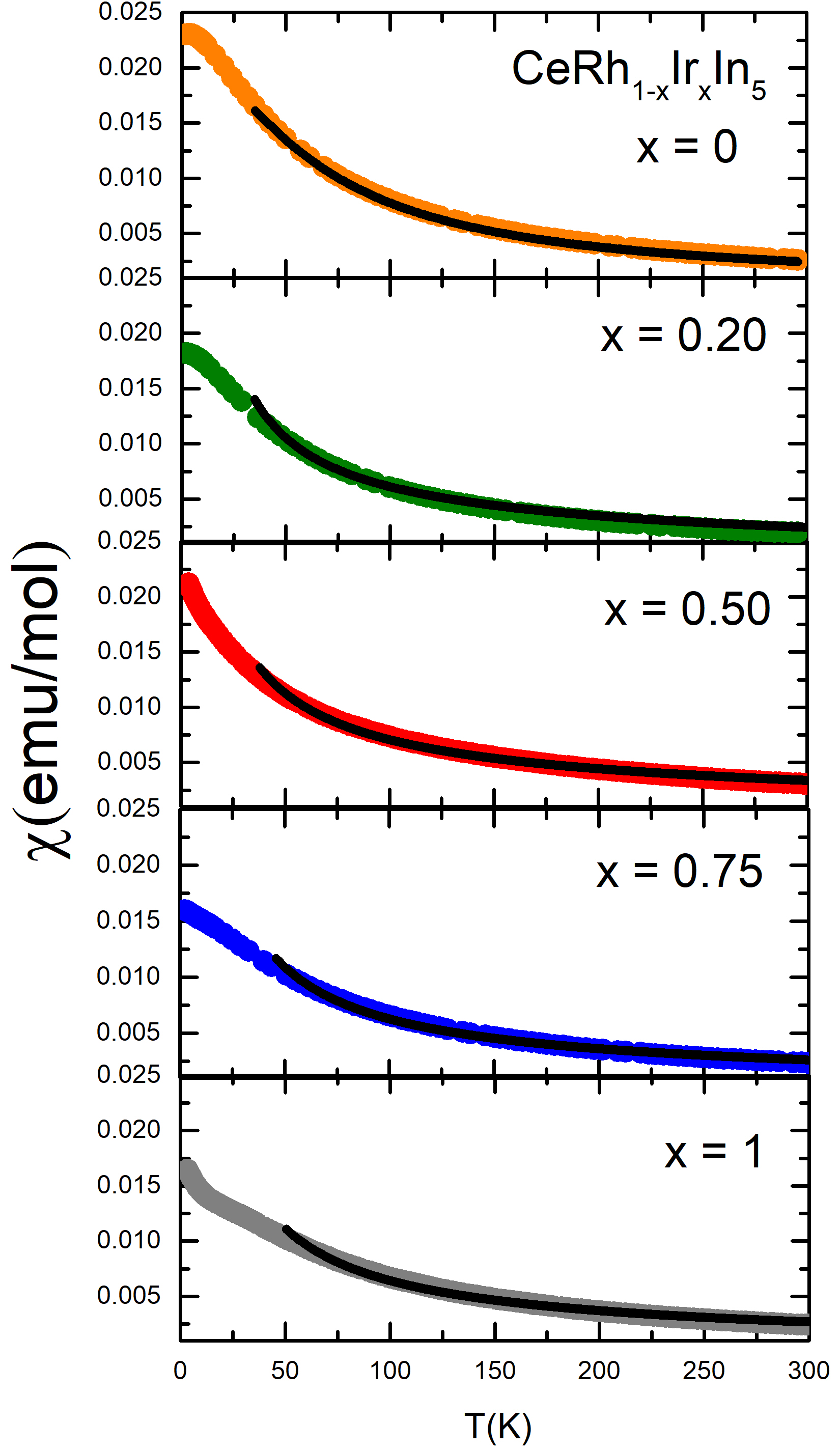}
\caption{\label{Fig:susceptibility} Bulk susceptibility along the $c$ axis of CeRh$_{1-x}$Ir$_x$In$_5$ versus temperature. The solid lines are fits to the high temperature data, as described in the text.}
\end{figure}

Figure \ref{Fig:susceptibility} shows the bulk magnetic susceptibility, $\chi$, of a series of single crystals of CeRh$_{1-x}$Ir$_x$In$_5$ between 4 and 300 K. At high temperature $\chi$ displays Curie-Weiss behavior, reflecting the localized nature of the Ce $4f$ electrons. For $T> 50$ K, $\chi$ is well-described by local moments in a tetragonal crystal field, with an effective exchange field, as discussed in \cite{Pagliuso2006} and shown as solid lines.   This behavior is modified at low temperature due to the crystal field splitting, the Kondo interaction, and the exchange interaction among the Ce orbitals, all of which depend on the doping, $x$. As  $x$ increases the ground state evolves from antiferromagnetic below $T_N = 3.8$ K ($x=0$) to superconducting below $T_c = 0.4$ K ($x=1$), with a possible quantum phase transition near $x=0.3$ \cite{RhIr115PRLKitaoka}. de Haas-van Alphen measurements and band structure calculations indicate that the 4$f$ electrons become more itinerant, and this trend is reflected in the overall decrease in the magnitude of $\chi$ over this range \cite{ShishidoRh115dHvA,Ir115dHvA}.

\subsection*{Knight shift}
\begin{figure}
\centering
\includegraphics[width=\linewidth]{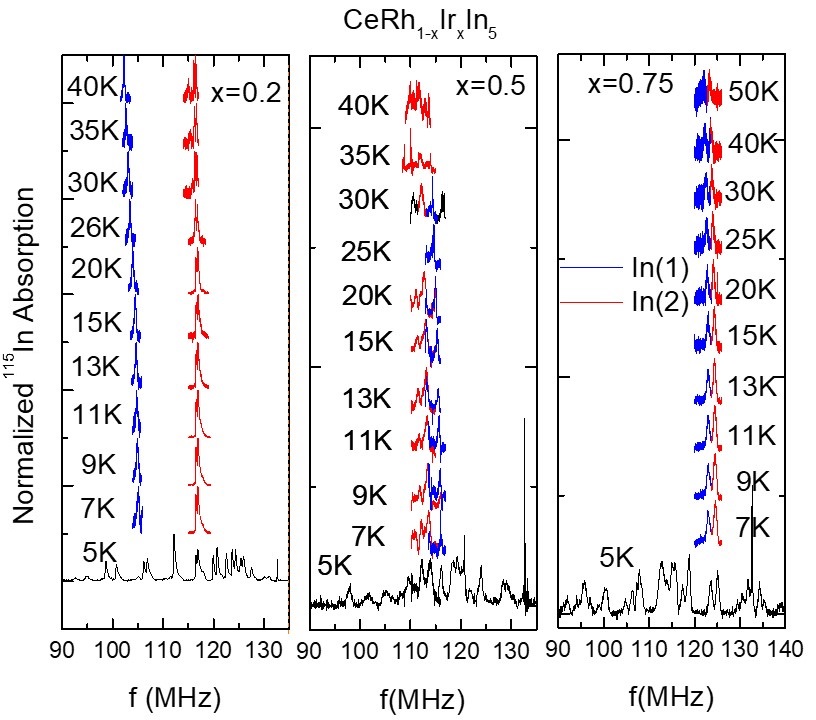}
\caption{\label{Fig:1} Spectra of CeRh$_{1-x}$Ir$_{x}$In$_5$ for $x=0.20$, $x=0.50$ and $x=0.75$ for several different temperatures for  $H_0 = 11.7$\,T along the $c$ direction. Blue indicates In(1) and red indicates In(2).  Slight misalignments  cause the In(2) spectra to split. Details of the spectral fitting are provided in the supplemental information.}
\end{figure}

\begin{figure}
\centering
\includegraphics[width=1.1\linewidth]{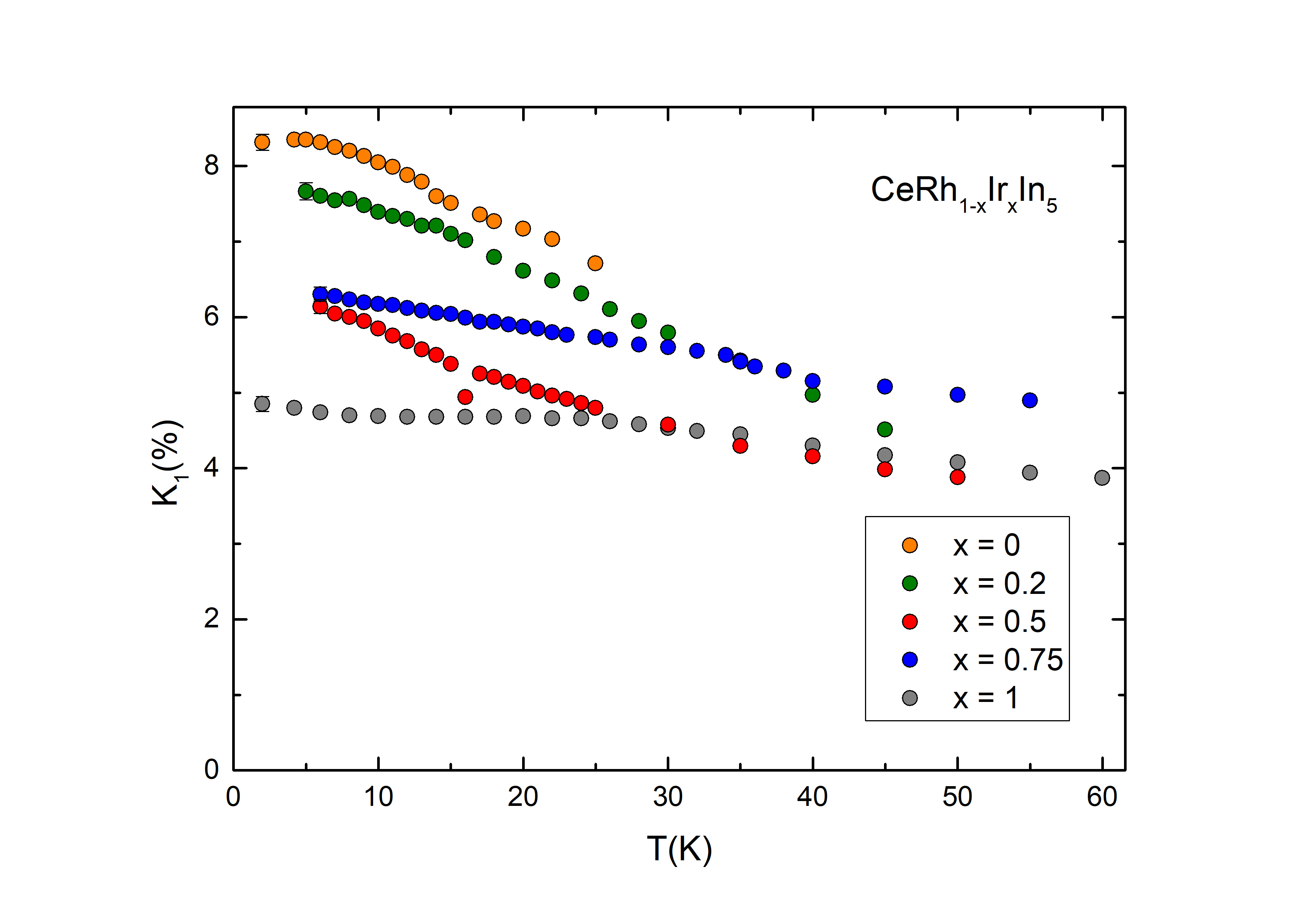}
\caption{\footnotesize In(1) Knight shift versus temperature, extracted from the spectra displayed in Fig. \ref{Fig:1}.  Data for the $x=0$ and $x=1$ are reproduced from \cite{ShirerPNAS2012}.}
\label{fig:K1}
\end{figure}

Figure \ref{Fig:1} shows $^{115}$In NMR spectra of CeRh$_{1-x}$Ir$_x$In$_5$ for several different values of $x$ in the field $H_0 = 11.7294$ T along the $c$-axis. In ($I=9/2$) has nine transitions for each site, and all transitions are split by the quadrupolar interaction \cite{Lin2015}. The resonance frequencies of the In(1) depend on the Knight shift, $K$, the field orientation, and the electric field gradient (EFG) along the c-axis, $\nu_{cc}$, as described in the supplemental information.  The In(2) resonances are more complex due to the non-zero EFG asymmetry parameter, $\eta = |\nu_{xx}|-|\nu_{yy}|)/|\nu_{zz}|$, and misorientations from $\mathbf{H}_0 || c$ split the In(2) resonances giving rise to double peaks. Furthermore, local disorder and mixing of the Rh and Ir in the doped samples are responsible for multiple In(2) sites and create complex NMR spectra.    The In(1) sites were identified by fitting the spectra with exact diagonalization \cite{Shockley2015}, and the results are summarized in Table \ref{tab:pars} and shown in the Supplemental Information. The EFG values agree with previous measurements \cite{RhIr115PRLKitaoka}.  Although we are able to fit the full In(1) spectra, there are several In(2) peaks that are not fully identified. The Knight shifts of the In(1) were determined as a function of temperature by measuring specific transitions and subtracting the quadrupolar shift. The resonances exhibit large shifts with temperature, consistent with Knight shifts observed in pure CeRhIn$_5$ and CeIrIn$_5$.   The In(1) shift, $K_1$, is shown in Fig. \ref{fig:K1}  as function of temperature and doping, $x$.   $K_1$ decreases with $x$, similar to $\chi$, although there are small deviations at low temperature.

\begin{table*}
\centering
\caption{\label{tab:pars} EFG, Hyperfine, and CEF Parameters for CeRh$_{1-x}$Ir$_x$In$_5$ and CeCoIn$_5$. Values for $x=0$, $x=1$, and CeCoIn$_5$ are reproduced from \cite{ShirerPNAS2012}, \cite{Curro2004}, and  \cite{WillersCEF115s}. Hyperfine couplings are given in units of kOe/$\mu_B$.}

\begin{ruledtabular}
\begin{tabular}{lddddddddd}
\textbf{x}&
\multicolumn{1}{c}{$\nu_{zz}(1)$ (MHz)}&
\multicolumn{1}{c}\textbf{$\nu_{zz}(2)$ (MHz)}&
\multicolumn{1}{c}\textbf{$\eta(2)$}&
\multicolumn{1}{c}\textbf{$B_{cc}(1)$ } &
\multicolumn{1}{c}\textbf{$B_{cc}(2)$ } &
\multicolumn{1}{c}\textbf{$b_{20}$ (meV)} &
\multicolumn{1}{c}\textbf{$b_{40}$ (meV)}  &
\multicolumn{1}{c}\textbf{$|b_{44}|$ (meV)}  &
\multicolumn{1}{c}\textbf{$\alpha^2$}\\
\colrule
0 & 6.78   & 16.665   & 0.445 & 21.4 &  8.6 & -0.928 & 0.052   & 0.128     & 0.407   \\
0.20   & 6.4(5)  & 17.3(8)   & 0.45 & 20.0(1.0) & --  &  -0.961  & 0.057   & 0.118  &  0.37   \\
0.50   & 6.3(5)   & 16.5(8)   & 0.45 & 16.7(2) &  --  & -0.996  & 0.061 & 0.107 &  0.28    \\
0.75    & 6.2(5)    & 17.3(8)   & 0.45 & 15.0(2.0) &  -- & -1.154   & 0.068   & 0.86 & 0.26      \\
1    & 6.07    & 18.17   & 0.46 & 13.8 &   15.9 & -1.197 & 0.069   & 0.088   &  0.250  \\
\colrule
\textrm{CeCoIn$_5$} & 8.173    & 15.489   & 0.386 & 8.9&  28.1 & -0.856 & 0.063   & 0.089   &  0.129
\end{tabular}
\end{ruledtabular}
\end{table*}

\section*{\label{sec:resultsanddiscussion}Discussion}

\subsection*{Magnetic Response of Ce Orbitals}  
\begin{figure}
\centering
\includegraphics[width=\linewidth]{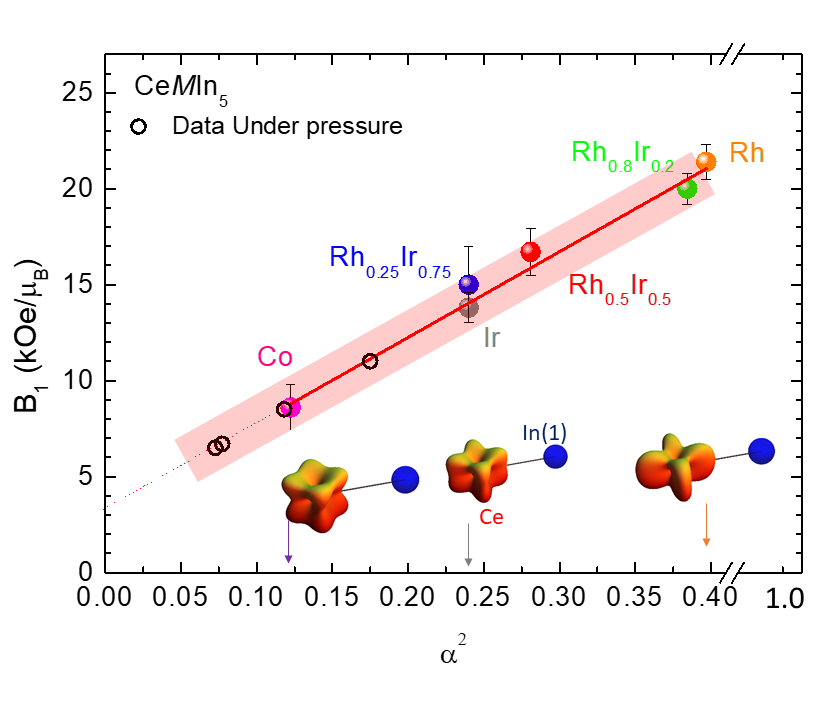}
\caption{\footnotesize The transferred hyperfine coupling to the In(1) as a function of $\alpha^2$.  The solid line is a fit to  $B_1(\alpha^2) = B_{10} + \kappa \alpha^2$, where $B_{10} = 3.3(8)$ kOe/$\mu_B$ and $\kappa = 44(2)$ kOe$\mu_B$. The open points are taken from Ref. \cite{Lin2015} and inferring the $\alpha^2$ values using the fit to the ambient-pressure data (shown in Fig. \ref{fig:pressure}).  The insets show how the spatial form of the Ce 4$f$ wavefunction evolves as $\alpha^2$ changes.}
\label{fig:B1vsalpha}
\end{figure}

The behavior of an isolated Ce 4f electron spin in a tetragonal environment is given by
\begin{equation}
    \mathcal{H}_{CEF} = b_{20} \hat{O}_2^0 + b_{40} \hat{O}_4^0 + b_{44} \hat{O}_4^4,
    \label{eq:CEF}
\end{equation}
where $\hat{O}_n^m$ are the Stevens operators and $b_{nm}$ are parameters that characterize the crystal field.  The $J=5/2$ multiplet is split into three Kramers' doublets: $\Gamma_7^{(1)}$ $\Gamma_7^{(2)}$, and $\Gamma_6$.  The ground state $\Gamma_7^{(1)}$ wavefunctions can be expressed as:
\begin{equation}
|\psi^{\Gamma_7}_{1,2}\rangle = \alpha\left|\pm\frac{5}{2}\right\rangle\pm\sqrt{1-\alpha^2}\left|\mp\frac{3}{2}\right\rangle,
\end{equation}
where $\alpha$  characterizes the degree of mixing between the $J_z$ manifolds and controls the degree of spatial anisotropy of the orbital.   $\alpha^2_c = 1/6$ for cubic symmetry ($b_{44}=4b_{40}$ and $b_{20} = 0$).  $\alpha^2$ increases as the CEF potential becomes more tetragonal, and the orbital shape becomes more two-dimensional, as illustrated in Fig. \ref{fig:B1vsalpha}.

The magnetic susceptibility of the $f$ moments is given by $\chi_{ff}^{-1} = \chi_{ff0}^{-1} + \lambda$, where $\lambda$ is a mean-field parameter that captures both the exchange and Kondo interactions, and $\chi_{0}$ is given by:
\begin{equation}
\chi_{ff0} = {N_A}\frac{\mu_B^2}{Z}\sum_{i,j}e^{-\beta E_i}\int_0^{\beta}\left|\langle i|g_J\hat{J}_z|j\rangle\right|^2 e^{(E_i-E_j)\tau}d\tau,
\end{equation}
where $N_A$ is Avogadro's number, and the sum is over the eigenstates of $\mathcal{H}_{CEF}$ (Eq. \ref{eq:CEF}). We assume that the f-moments dominate the magnetic response and fit the susceptibility data using the parameters $b_{20}$, $b_{40}$, $b_{44}$  and $\lambda$, as shown in Fig. \ref{Fig:susceptibility} and using the values shown in Table \ref{tab:pars}.  Although fitting the susceptibility data is an indirect and less-precise approach to extracting the CEF parameters than direct XAS measurements, our values agree well with published results \cite{Severing}.

\subsection*{Hyperfine Couplings} 
\begin{figure}
\centering
\includegraphics[width=\linewidth]{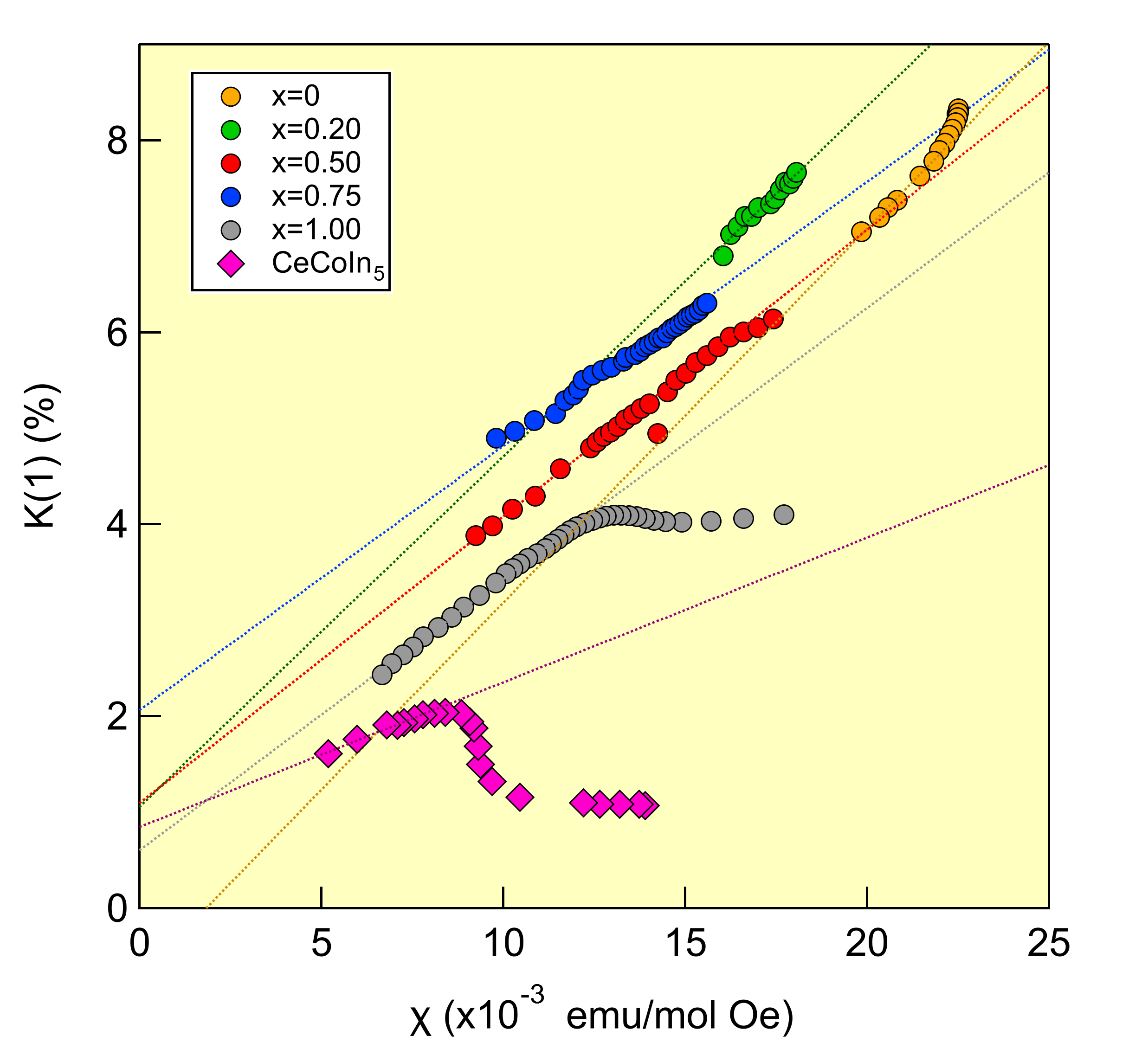}
\caption{\footnotesize In(1) Knight shift versus magnetic susceptibility. Dotted lines are fits to the high temperature ($T>T^*$) data. The slopes are reported in Table \ref{tab:pars}. Data for CeCoIn$_5$ is reproduced from \cite{Curro2001}.}
\label{Fig3}
\end{figure}

For the hyperfine interaction in heavy fermions, the Knight shift is given by $K = A\chi_{cc} + (A+B)\chi_{cf} + B\chi_{ff}$,  where $\chi_{cc}$, $\chi_{ff}$, and $\chi_{cf}$  are contributions from the conduction electrons, the $f$-moments, and their interaction, respectively \cite{Curro2006a}.  Here we write $B = nB_{cc}$, where $n$ is the number of nearest neighbor $f$ site ($n=4$ for In(1) and 2 for In(2)), and $B_{cc}$ corresponds to the $c$-axis component of the tensor $\mathbb{B}$. The total susceptibility is $\chi = \chi_{cc} + 2\chi_{cf} + \chi_{ff}$. For sufficiently high temperatures $T>T^*$ the first two contributions to the shift can be ignored and  $K\approx B\chi$. A plot of $K_1$ versus $\chi$ yields a straight line for $T>T^*$ with slope $B_1$, as shown in Fig. \ref{Fig3}.  We fit this data to extract the transferred hyperfine couplings as shown in Table \ref{tab:pars}.  There are two important trends evident in the data.  First, the slope decreases with increasing $x$, and gets even smaller in CeCoIn$_5$.  Secondly, there is a breakdown in the linear relationship below a temperature $T^*$.  This deviation signals the onset of coherence at low temperatures, where $\chi_{cc}$ and $\chi_{cf}$ can no longer be ignored.  As $x$ increases from 0 to 1, the deviation evolves from curving upwards (pure CeRhIn$_5$) to downwards (undoped CeIrIn$_5$), and becomes more pronounced for CeCoIn$_5$.  For intermediate values of $x \approx 0.5$, there is no clear evidence of such a deviation in the data.  This behavior reflects changes in the relative size of the on-site $A$ coupling and the transferred $B$ coupling, as well as the growth of heavy fermion coherence, as discussed below.

We can now examine how the hyperfine coupling correlates with the CEF ground state orbital shape. Fig. \ref{fig:B1vsalpha} shows $B_1$ versus $\alpha^2$ using the parameters extracted from the fits to the Knight shift and susceptibility. The solid line in Fig. \ref{fig:B1vsalpha} is a linear fit to the data, including the value for CeCoIn$_5$.   As $\alpha^2$ increases, the orbitals become more oblate, the lobes pointing along the In(1) directions become more extended, and the transferred hyperfine coupling to the In(1) nucleus increases. The amplitude of the 4f wavefunction along the Ce-In(1) bond direction varies as $1 + 4\alpha^2 + 2\sqrt{5 \alpha^2(1-\alpha^2)}$, which is approximately linear over this range of $\alpha^2$, in agreement with our observation. This result reveals directly how Ce-In(1) hybridization evolves as the ground state orbital anisotropy changes.

\begin{figure}
\centering
\includegraphics[width=\linewidth]{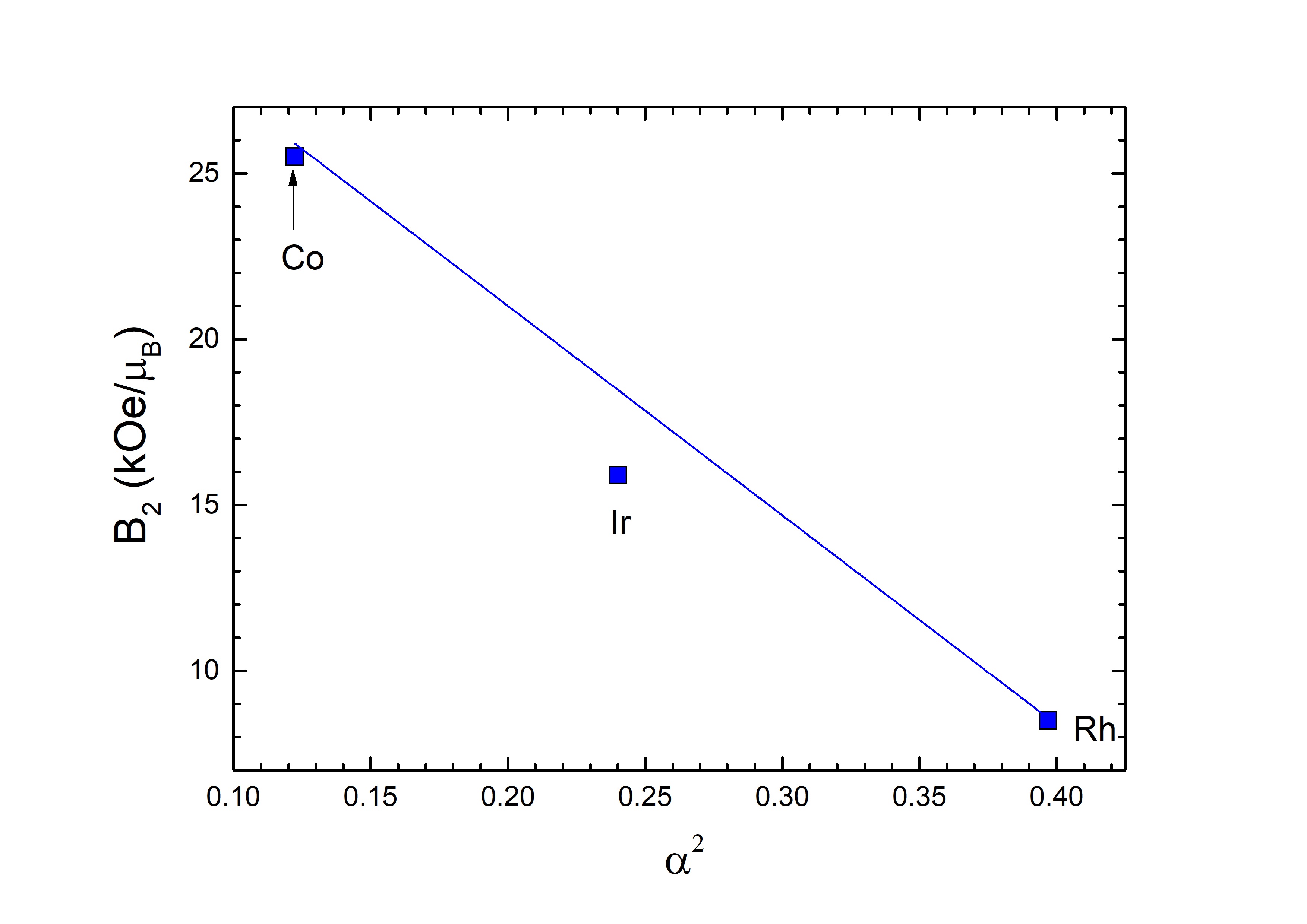}
\caption{\footnotesize Hyperfine coupling to the In(2) as a function of $\alpha^2$. The solid line is a guide to the eye.}
\label{fig:In2vsalpha}
\end{figure}

Direct measurements of the In(2) hyperfine coupling are challenging in the doped materials.  However, it is insightful to plot the values for the pure CeRhIn$_5$, CeIrIn$_5$ and CeCoIn$_5$ versus $\alpha^2$, as shown in Fig. \ref{fig:In2vsalpha}.  In contrast to the In(1), $B_2$ is largest when the orbital lobes extend out of the plane and decreases with increasing $\alpha^2$.  In this case, the amplitude of the 4f wavefunction along the Ce-In(2) bond direction varies approximately as $51 - 6 \alpha (6 \alpha + \sqrt{5 - 5 \alpha^2})$, depending on the exact position of the In(2) in the unit cell.  This quantity decreases approximately linearly with $\alpha^2$, as observed in Fig. \ref{fig:In2vsalpha}. These results clearly indicate that as the in-plane hybridization to the In(1) site increases, the out-of-plane hybridization to the In(2) decreases.

\begin{figure}
\centering
\includegraphics[width=\linewidth]{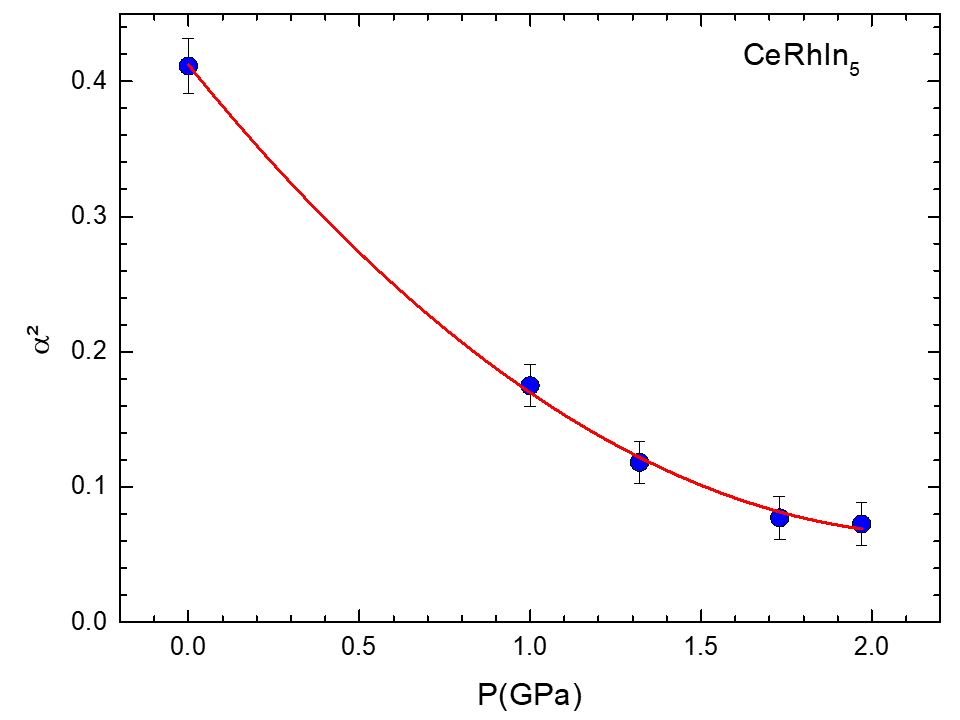}
\caption{\footnotesize $\alpha^2$ inferred from the pressure dependence of $B_1$ in \cite{Lin2015} using the relationship measured at ambient pressure in Fig. \ref{fig:B1vsalpha}. The solid line is a guide to the eye.}
\label{fig:pressure}
\end{figure}

Similar changes in the transferred hyperfine couplings were observed in CeRhIn$_5$ under hydrostatic pressure \cite{Lin2015}. In this case, $B_1$ decreases from 25.6 to 5.2 kOe/$\mu_B$ between ambient pressure and $\sim$ 2 GPa, which is close to the value observed in pure CeCoIn$_5$.  Using the linear relationship between $B_1$ and $\alpha^2$ we observe under ambient pressure in Fig. \ref{fig:B1vsalpha}, we infer the pressure dependence of $\alpha^2$ in Fig. \ref{fig:pressure}.  These results imply that pressure changes the CEF parameters so that the Ce wavefunction lobes extend more out of the plane, becoming more cubic-like.   This observation is consistent with the fact that at this pressure, CeRhIn$_5$ becomes superconducting with $T_c$ similar to CeCoIn$_5$, and develops a large Fermi surface \cite{tuson,ShishidoRh115dHvA}.  On the other hand, the data imply that the $\alpha^2$ becomes smaller than $\alpha_c=1/6$ which may be unphysical.  It should be noted, however, the values for $B_1$  in \cite{Lin2015} were determined based on the assumption that the In(2) hyperfine coupling did not change under pressure.  Because there is no independent susceptibility data under pressure, it was only possible to directly extract the ratio of $B_1(P)$ to $B_2(P)$, rather than their independent values.  Nevertheless, the trend under hydrostatic pressure is similar to that observed with `chemical pressure'.

\subsection*{Evolution of Heavy Fermion Coherence}
\begin{figure}
\centering
\includegraphics[width=\linewidth]{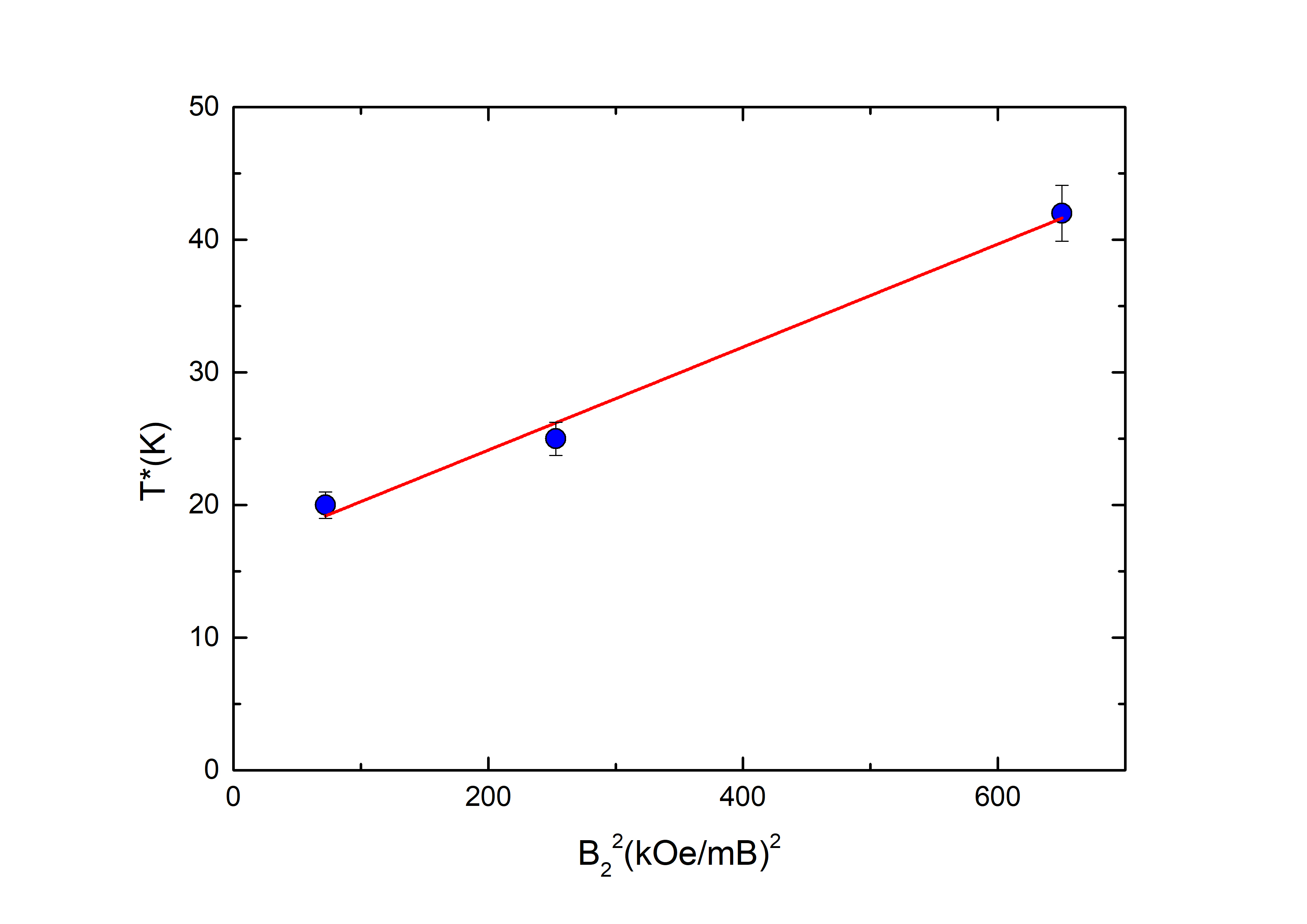}
\caption{\footnotesize The coherence temperature $T^*$ versus $B_2^2$. The solid line is given by $T^*=16.3(2) + 0.039(3)B_2^2$.}
\label{fig:Tstar}
\end{figure}

The Knight shift anomaly observed in Fig. \ref{Fig3} reflects the growth of $K_{HF}(T)$ below $T^*$, where $K_{HF}(T) \propto (A-B)(1-T/T^*)^{3/2} [1 + \ln(T^*/T)]$ \cite{ShirerPNAS2012}.   As $B_1$ increases with $\alpha^2$, the sign of $K_{HF}(T)$ changes from positive to negative, and vanishes when the transferred coupling equals the on-site coupling, $A$.  Previously, $A\approx 14$ kOe/$\mu_B$ was estimated in CeCoIn$_5$ \cite{Curro2004}.   Since this quantity reflects a combination of a Fermi-contact interaction plus core-polarization from the Indium 5p orbitals, it should only weakly depend on doping or transition metal element. Thus, the observation that the Knight shift anomaly is small or absent in CeRh$_{1-x}$Ir$_x$In$_5$ for $x\approx 0.5$ in Fig. \ref{Fig3}  likely reflects the fact that $B_1 \approx A$ in this range.  On the other hand, there is a clear growth of both the magnitude of $K_{HF}$ and the onset temperature, $T^*$, as $\alpha^2$ decreases in CeIrIn$_5$ and CeCoIn$_5$. $T^*$ is approximately the temperature where the entropy reaches $R\ln 2$, and has been shown empirically related to the Kondo coupling, $J$, as $T^* =0.45 J^2\rho$, where $\rho$ is the density of conduction electron states \cite{YangPinesNature}. Since the Kondo coupling arises due to hybridization of the $4f$ orbital, it is natural to expect that $J\propto\alpha^2$, and thus $T^* \sim B_2^2$, as demonstrated in Fig. \ref{fig:Tstar}. This observation supports previous dynamical mean-field theory (DMFT) calculations of CeIrIn$_5$ that indicated the hybridization gap for the Ce-In(2) band dominates that for the in-plane Ce-In(1) band \cite{shim2007modeling}.  The Kondo hybridization, which drives the low temperature correlated behavior, is thus directly controlled by the shape of the Ce 4f orbitals.

\section*{\label{sec:conclusion}Summary}
In summary, we have measured the Knight shift and magnetic susceptibilities for several single crystals of CeRh$_{1-x}$Ir$_x$In$_5$, and found that the transferred hyperfine coupling to the In(1) is suppressed by nearly a factor of two as $x$ varies from 0 to 1.  We extract the crystal field parameters by fitting the bulk magnetic susceptibility and find values that agree well with recent XAS results.  Our results reveal that the hyperfine coupling varies linearly with  $\alpha^2$, where $\alpha$ describes the admixture between the $|J_z=\frac{3}{2}\rangle$ and $|J_z=\frac{5}{2}\rangle$ states of the Ce 4f ground state wavefunction. The hyperfine coupling to the In(2) exhibits the opposite trend, and is enhanced as the wavefunction extends out of the plane. These results provide direct proof that the hybridization between the 4f wavefunction and the In 5p orbitals is controlled by the orbital anisotropy.  This observation also offers an explanation for the abrupt decrease in the In(1) Knight shift at 30 T, where the magnetic field induces changes to the crystal field ground state orbital \cite{Lesseux2020}.  This study establishes transferred hyperfine interactions as an important probe of hybridization anisotropy in heavy fermion materials, and may provide a more straightforward approach to determine the crystal field parameters.  Further experiments exploring  all components of the hyperfine coupling tensor will be important to better characterize this hybridization, especially under pressure.

\section*{Methods}

\noindent \textbf{Samples}:
High-quality single crystals of CeRh$_x$Ir$_{1-x}$In$_5$ with multiple values of doping, $x$, were grown via the flux method technique \cite{Zach}, in which stoichiometric quantities of Ce, Rh, and Ir were located in a ceramic crucible with excess of In (20 mols, instead of 5). The crucible was sealed within a quartz tube and heated in a furnace following the temperature ramp reported on \cite{Pagliuso_2001}. Single crystals of tetragonal shape were selected manually and polished to remove In flux from the surface. Powder x-ray diffraction measurements in a Bruker Phaser D2 diffractometer with Cu K$_\alpha$ radiation ($\lambda$ = 1.5418 \AA) using a silicon plate with zero background confirmed single phase purity for these samples. The diffraction pattern confirmed the expected  tetragonal phase without evidence of spurious phases. \\

\noindent \textbf{Magnetic Susceptibility Measurements}:
Magnetic susceptibility experiments were carried out on a commercial Quantum Design PPMS-14T, with a insert for VSM magnetization measurements in the range 3 $<$ $T$ $<$ 300\,K, in a magnetic field of 8\,T.\\

\noindent \textbf{NMR Measurements}:
NMR experiments were carried out in a Oxford 11.7\,T high-homogeneity fixed-field magnet, equipped with a Janis closed-cycle cryostat in the temperature range of 4 $<$ $T$ $<$ 300\,K, and data were collected using a TecMag Apollo spectrometer. Samples were mounted with $\mathbf{H}_0 || c$, and radiofrequency pulses of varying duration (1.9 -2.2\,$\mu$s) were used. Spectra at multiple frequencies were acquired and summed over a broad range to identify multiple satellites of each In site, as shown in the supplementary information.\\

\acknowledgments{The authors gratefully acknowledge support from the UC Davis Seed Grant program and FAPESP (Grants No. 2016/14436-3, 2018/11364-7, 2017/10581-1 and 2012/05903-6). RRU acknowledges CNPq Grant No 309483/2018-2. PGP acknowledges CNPq Grant No 304496/2017-0. We acknowledge helpful discussions with T. Kissikov and P. Klavins. Work at UC Davis was partially supported by the NSF under Grant No. DMR-1807889.}

\bibliography{CeRhIrIn5references}

\newpage
\section*{Supplemental Information}

\subsection*{NMR Spectral Fitting}
In order to identify the indium sites, the NMR spectra for CeRh$_{1-x}$Ir$_x$In$_5$ at 5K were fit using the nuclear spin Hamiltonian:
\begin{equation}
  \mathcal{H} = \gamma\hbar{\mathbf{I}}\cdot(1+\mathbf{K})\cdot \mathbf{H}_0 + \frac{h\nu_{zz}}{6}[3{I}_z^2-\hat{I}^2 + \eta\left(\hat{I}_x^2-{I}_y^2\right)]
  \label{eqn:Hnuc}
\end{equation}
where $\gamma=0.93295$ kHz/G is the gyromagnetic ratio,  ${I}_{\alpha}$ are the nuclear spin operators, $\mathbf{K}$ is the Knight shift tensor, $\nu_{zz}$ is the largest eigenvalue of the EFG tensor, and $\eta$ is the asymmetry parameter.  The $(x,y,z)$ coordinates are defined in the usual manner such that $|\nu_{zz}|>|\nu_{xx}|> |\nu_{yy}|$.  For the In(1), $z$ corresponds to the $c$ axis, and for the In(2), $z$ corresponds to the direction normal to unit cell face containing the In(2) atom, whereas $x$ corresponds to the other in-plane direction perpendicular to $c$.  To fit the spectra, we assume the direction of $\mathbf{H}_0$ is described by the spherical polar angles $(\theta,\phi)$ with respect to the $c$ axis, and perform an exact diagonalization of \ref{eqn:Hnuc}.  The peaks are fit to Voigt functions. For the case where $\theta\neq 0^{\circ}$ the In(2) sites split into two peaks.  We find that there are, in fact, more In(2) sites that remain unidentified that may be associated with local disorder due to the non-stoichiometry of Rh and Ir.

\begin{figure*}[h]
\centering
\includegraphics[width=\textwidth]{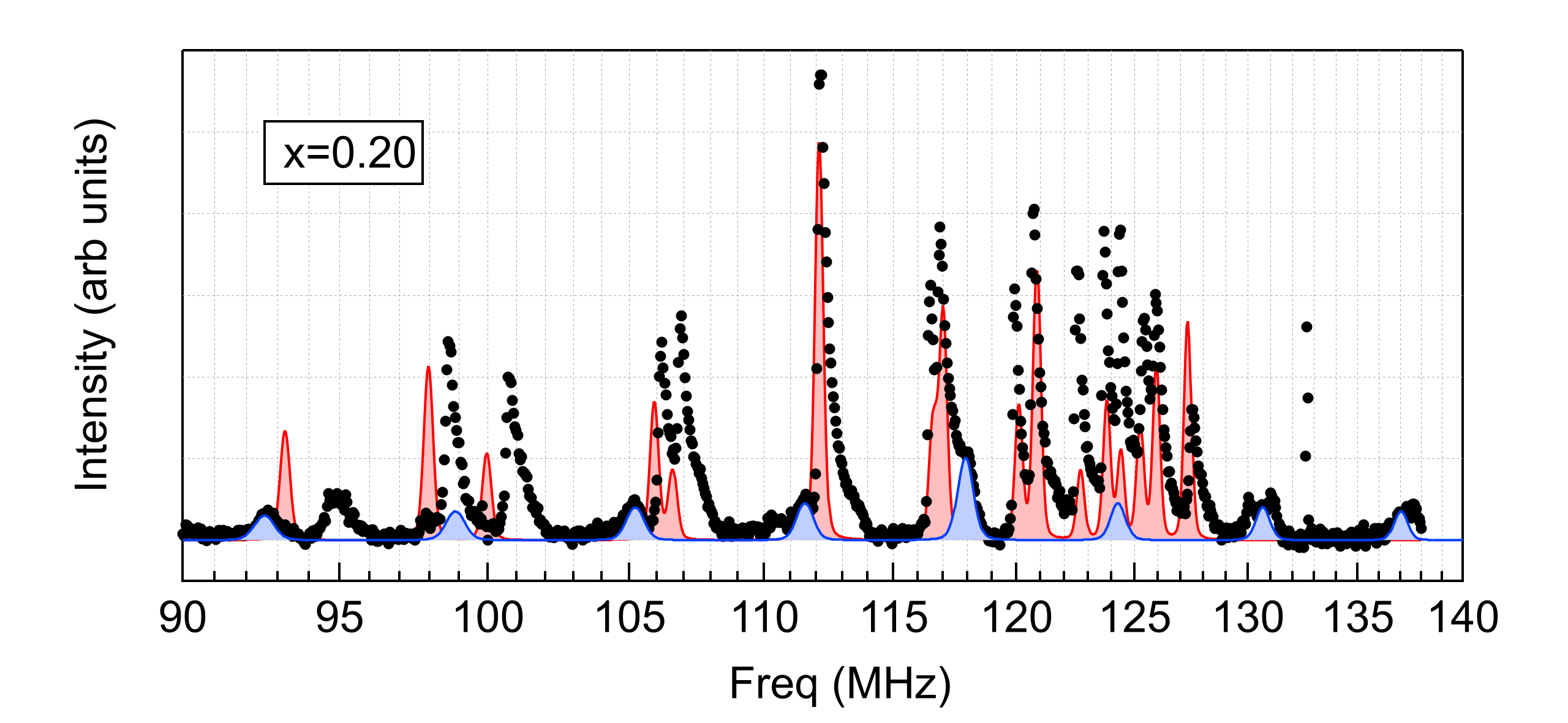}
\caption{\footnotesize NMR spectrum for CeRh$_{1-x}$Ir$_x$In$_5$ with $x=0.20$. Fits to the In(1) sites are shown in blue and those for the In(2) sites are shown in red.  The fit parameters are: $\theta = 4\pm 3^{\circ}$, $\phi = 8\pm 3^{\circ}$, $K_c(1) = 7.8\pm 0.1\%$, $K_c(2) = 2.5\pm 0.1\%$, $\nu_{zz}(1) = 6.41\pm 0.5$ MHz, $\nu_{zz}(2)=17.3\pm 0.5$ MHz, and $\eta(2) = 0.45\pm 0.02$. }
\label{fig:fit20}
\end{figure*}

\begin{figure*}[h]
\centering
\includegraphics[width=\textwidth]{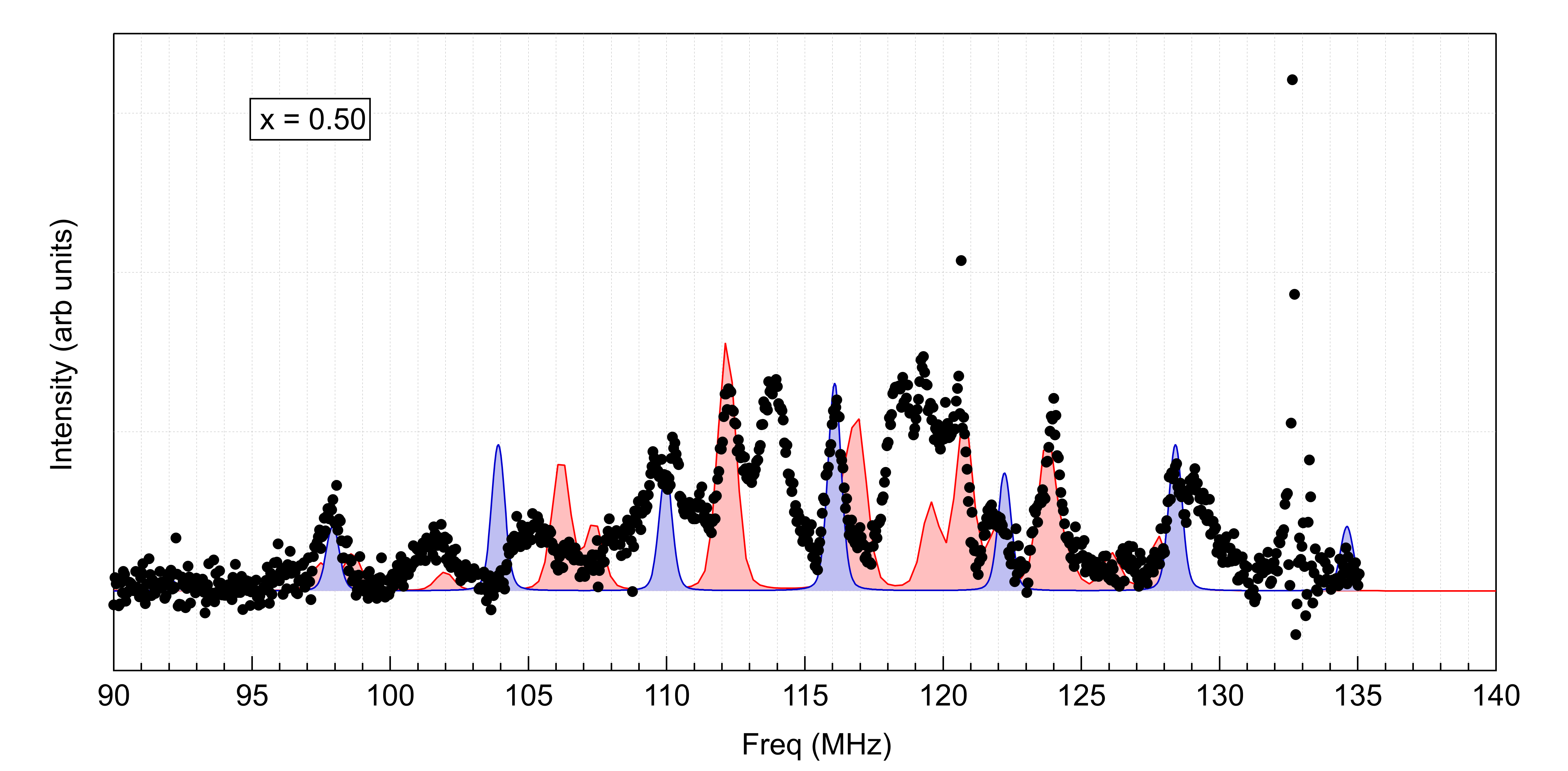}
\caption{\footnotesize NMR spectrum for CeRh$_{1-x}$Ir$_x$In$_5$ with $x=0.50$. Fits to the In(1) sites are shown in blue and those for the In(2) sites are shown in red.  The fit parameters are: $\theta = 8.3\pm 3^{\circ}$, $\phi = 11\pm3^{\circ}$, $K_c(1) = 6.2\pm 0.2\%$, $K_c(2) =  2.8\pm 0.5\%$, $\nu_{zz}(1) = 6.32\pm 0.2$ MHz, $\nu_{zz}(2)= 16.57\pm 0.5$ MHz, and $\eta(2) = 0.45\pm0.02$. }
\label{fig:fit50}
\end{figure*}

\begin{figure*}[h]
\centering
\includegraphics[width=\textwidth]{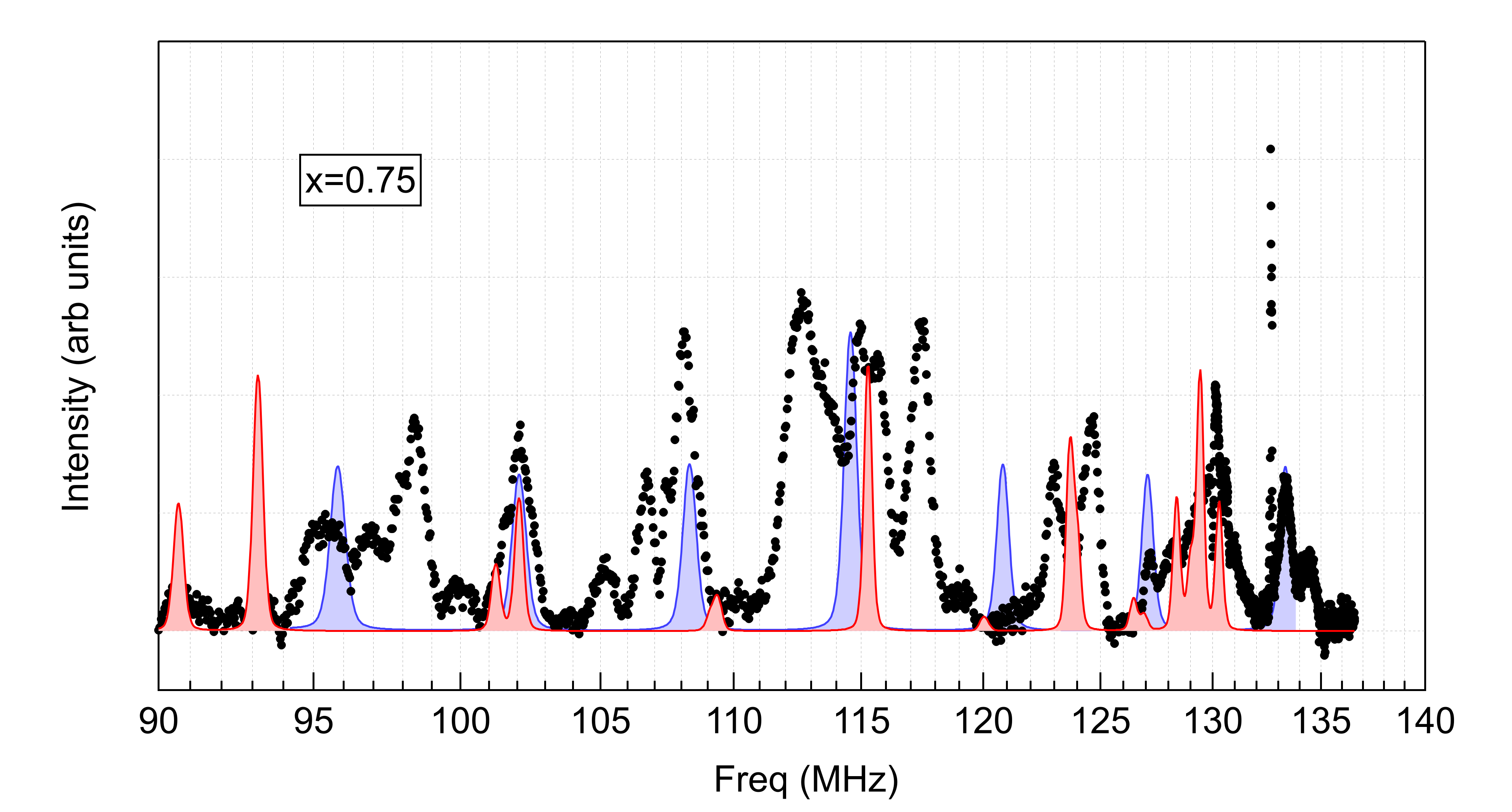}
\caption{\footnotesize NMR spectrum for CeRh$_{1-x}$Ir$_x$In$_5$ with $x=0.75$. Fits to the In(1) sites are shown in blue and those for the In(2) sites are shown in red.  The fit parameters are: $\theta = 4\pm 3^{\circ}$, $\phi = 0^{\circ}$, $K_c(1) = 4.7\pm 0.2\%$, $K_c(2) = 5.5\pm 0.8\%$, $\nu_{zz}(1) = 6.25$ MHz, $\nu_{zz}(2)= 17.28\pm 0.5$ MHz, and $\eta(2) = 0.45\pm0.02$. }
\label{fig:fit75}
\end{figure*}

\end{document}